\documentclass[pra,twocolumn,showpacs,groupedaddress]{revtex4}

\usepackage{amsmath}
\usepackage{amsfonts}
\usepackage{stmaryrd}
\usepackage{amssymb}
\usepackage{graphics,epsfig}
\usepackage{bm}

\begin{document}

\title{Superposition and Entanglement from Quantum Scope}

\author{Dong-Sheng Wang}
\email{wdsn1987@gmail.com}

\date{26 September 2011}

\begin{abstract}
The abstract framework of quantum mechanics (QM) causes the
well-known weirdness, which leads to the field of foundation of QM.
We constructed the new concept, i.e., {\em scope}, to lay the
foundation of quantum coherence and openness, also the principles of
superposition and entanglement. We studied analytically and
quantitatively the quantum correlations and information, also we
discussed the physical essence of the existed entanglement measures.
We compared with several other approaches to the foundation of QM,
and we stated that the concept of scope is unique and has not been
demonstrated before.
\end{abstract}


\maketitle

\section{INTRODUCTION}
\label{sec:INTRODUCTION}

Quantum mechanics (QM), particularly featured by $\hbar$ and
$|\psi\rangle$, has been always facing the fundamental critics since
its establishment \cite{Bohm,Jammer,bell87}. The exact physical
meaning of ``quantum'' is unclear, which leads to the well-known
``weirdness''. The development of quantum field theory (QFT), which
is often viewed as an improved form of QM, does not resolve the
seminal problems in QM, such as the direct physical meaning of wave
function and phase, the role of quantization and measurement, etc.
The problem of the foundation of QM has been widely concerned again
due to the progress of quantum information and quantum computing
(QIQC) \cite{Nielsen,zurek03,schlosshauer,Timpson}. The concepts of
nonlocality and entanglement have proven their importance, yet,
there exist confusions between them \cite{NE}.

For the research of foundation of QM, we can briefly clarify two
sub-fields, one is the interpretation of QM
\cite{epr,Schrodinger,bohm52,neumann,everett,zeh,ballentine,pct,ch,cramer,fraassen,mermin,Dewdney,Aharonov,Aerts,Ronde,Jansson},
the other is the post-QM
\cite{Hardy,Fuchs,popescu,Bub,Gorobey,Butterfield,Ashtekar,Iliev,Isham07}.
The problems and confusions on basic concepts even philosophy, such
as the classical \cite{zurek03,schlosshauer}, hidden variable
\cite{bohm52}, collapse \cite{pct}, etc are mostly addressed in the
interpretations. Post-QM also devotes to the complete mathematical
form of QM, such as the information-theoretic approach \cite{Bub},
the K\"ahler structure approach \cite{Ashtekar}, etc.

The standard QM bases on several assumptions, and the physical
explanations of the Copenhagen (orthodox) interpretation are not
satisfying. Uncertainty and complementarity are emphasized; however,
for uncertainty, there are different explanations, either based on
measurement, knowledge, propensity, or reality \cite{Jammer}. For
the statistical meaning of the wave function and measurement, the
disagreement is more notable. Born viewed statistics as inherent,
while Einstein, by EPR-argument \cite{epr}, viewed statistics as a
result of incompleteness of QM. The notable ``jump'' and
``collapse'' do not have clear physical pictures. The most primary
problem is that the Copenhagen interpretation did not give an exact
and direct meaning to the wave function, e.g., why $|\psi\rangle$ is
complex?

Indeed, there are direct efforts to explain the physical meaning of
wave function, which led to the hidden variable method and Bohm
mechanics \cite{bohm52}. Every particle is endowed with definite
coordinate and trajectory under the restraint of guiding wave. Also,
there exists the physical collapse theory aimed to explain collapse
\cite{pct}. However, it modifies the Schr\"odinger equation and the
linearity, which is shown to be the analogy of the approaches based
on decoherence \cite{schlosshauer}. Coherence and decoherence, which
were not demonstrated in the standard QM, have gained lots of
attentions these years. For instance, the many-world interpretation
\cite{everett} views the universe as a coherent entity, and there
exist the inner observer and relative state. The consistent
(decohered) history approaches \cite{ch} take the time evolution and
measurement into account, and demonstrate the structure of quantum
logic. In Zurek's exsistential interpretation \cite{zurek03}, the
uncertainty principle is re-explained from the view of information,
and the Born's rule is deduced via the symmetry of envariance.
Recently, kinds of information-theoretic theories \cite{Bub} are
developed, where QM is viewed as a kind of information processing
theory. With all these progresses, however, the exact essence of
quantum has not been drawn.

At the same time, the continuous argument on foundation of QM also
arose the reflection of the primary philosophical conception, which
cannot be avoided in the quantum physics. The problem that why we
cannot easily understand QM just brings the problem that why we
think the classical mechanics (CM) is normal. In this paper, our
standing point is to view QM as a special kind of ``description'' of
motion, from which, there is no good or bad of QM and CM
\cite{bohr}. The theories, with equations and models, deal with the
same nature only revealing different aspects and properties. The
state vector in Hilbert space can capture the properties of
coherence, e.g., interference, which is exotic for CM. On the
contrary, trajectory is the basic idea in CM and the macroscopic
world, yet, not in QM, due to the uncertainty. We note the
trajectory in the path-integral method, which borrows ideas from CM,
only has mathematical meaning. In practice (reality), to employ
which description depends on its efficiency.

In this work, we start from one new concept, ``scope'', to check the
foundation of QM. We focus on the basic concepts and ideas in QM,
mostly relating to superposition. We do not intend to construct a
broader theory with the standard QM as the special case, the methods
based on scope mainly belong to the interpretation of QM. Scope
describes the ``structure of motion'', i.e., the logic and
systematic potential action region of the movement of a certain
object (system). That is, from scope we can get the ability and the
whole structure of the motion of the matter. The merits of the
concept of scope are multi-folds. Firstly, the direct one, the
weirdness of QM can be resolved based on scope, the wave function
and superposition can get their physical meaning. Secondly, the
abstract QM can gain solid natural foundation and physical essence,
i.e., QM is a totally new kind of description different with CM
(including statistical mechanics) and also Relativity. Thirdly,
scope can bring new ideas. For instance, the wave function and
superposition principle can also be used in the mesoscopic and
macroscopic scales, that is, the concept of scope is universal,
cannot be restricted by the scale. Also, QM may not rely on the
Hilbert space, since scope itself can form a kind of space, also
there can be other kinds of space, e.g., tangnet \cite{wang},
relating to quantum information and entanglement. Forth, another
point, the concept of scope may have new indication of the basic
ideas of nature and philosophy, e.g., it means every object has
finite ability of motion.

This work is divided into four parts. In Sec. \ref{sec:OPEN}, we
start from the concept of openness, which is central for QM, to
clarify several basic physical ideas. In Sec. \ref{sec:SCOPE}, we
introduce the new concept of ``scope'', from which we discuss the
physical meaning of superposition and entanglement, and to address
the foundation of QM. In Sec. \ref{sec:ENTANGLEMENT}, we study
several kinds of quantum states, and we analyze the degree of
entanglement and information. Several topics are discussed in the
appendix. One is the differences between scope and other existed
methods, where we state that the concept of scope has never been
proposed before. Another one is the model for the reduced
entanglement, by comparison with the two-body problem in CM. Last,
we study the physical meaning of entanglement measures at present,
like negativity, relative entropy of entanglement. In Sec.
\ref{sec:CONCLUSION}, we conclude and discuss briefly the physical
roles of scope in the foundation of QM.

\section{OPENNESS}
\label{sec:OPEN}

In this section, before the study on the concept of scope and
entanglement, we show that the concept of openness is fundamental in
QM. Openness means that the quantum object cannot be separated from
the outside world, and the basic subject in QM is the open system,
just the opposite of CM \cite{zeh,wang}. The physical reason for
openness is the existence of quantum coherence. We show that all the
basic equations in QM demonstrate the openness and coherence.

\subsection{Coherence} \label{sec:COHERENCE}

The quantum coherence leads to many crucial facts, such as the
double-slit interference, coherent and squeezed state of light, also
the Heisenberg's uncertainty principle, etc. Relating to
measurement, the uncertainty principle states that any measurement
disturbs the state of the quantum system, since the coherence is
disturbed. On the contrary, the nature light from sun is not
coherent as it is a kind of mixture. In CM, there is no coherence,
which is the main distinction between quantum and classical dynamics
\cite{wang}. Usually, in QM the method of density matrix \cite{Fano}
is employed to capture coherence, which is represented by the
non-diagonal elements. A diagonal density matrix is said to be
classical as the coherence vanishes. In QFT, there exists coherence
since the quantum field itself is the coherent entity, such as
vacuum, electromagnetical field etc, although the density matrix is
not often used. ``Coherent'' means that the field as a whole has
fixed amplitude and phase. The micro-particles, as the excited state
of field, have coherence since interactions between them are due to
the field. The facts of particle creation, classicality, and
decoherence lead to the degeneracy of coherence, yet, particle can
go back to its ground state via the annihilation operator, thus,
re-excite the coherence \cite{Calzetta,Habib}.

In addition, we note that the standard QM does not focus on the
properties of field, e.g., the creation and vanish of particles,
also the origin of ensemble, which we will discuss below.

\subsection{Ensemble} \label{sec:ENSEMBLE}

In QM, there was the argument that whether quantum theory describes
the dynamics of the single system or the ensemble \cite{ballentine}.
According to the standard QM, the existence of ensemble is {\em a
priori} fact. For instance, in the double-slit interference
experiment, the ensemble of electrons is used, the electrons are
viewed as identical, and the state of electrons are the
superposition of two modes corresponding to the two slits. Here, QM
cannot explain the origin of the ensemble; in contrast, QFT can give
the creation-annihilation picture of the ensemble of electrons. The
quantum states of each electron before entering the slit are not
necessarily orthogonal, and there exists field among the electrons,
that is, the existence of ensemble does not mean there is no
coherence.

Yet, the methods of QM are believed as universal \cite{everett},
thus, it should provide the physical picture for ensemble. Recently,
there are efforts to indicate the quantum origin of ensemble via
entanglement \cite{Popescu,Kim}. It is shown that the system within
the coherent universe behaves as the canonical ensemble, as if the
universe were in the mixed state, in which each pure state has the
equal probability. Physically, there can be kinds of coherence and
decoherence processes in QM, which relates to the distinction
between the fine-grained and coarse-grained description of states of
the system \cite{Calzetta}. We will further analyze this in the
study of density matrix below. In addition, we note that the concept
of ensemble is similar to ``mixture'', which is also widely employed
in QM. However, the ``mixed state'' is not an exact expression,
e.g., it does not indicate the state is of one single object or many
objects. We view mixture as the classical concept.

\subsection{Openness} \label{sec:OPENNESS}

The coherence indicates the openness of quantum dynamics. Here we
show that the basic equations of QM all demonstrate the openness.
The Schr\"odinger equation $i\hbar |\dot{\Psi}\rangle=\mathcal
{H}|\Psi\rangle$, where $|\Psi\rangle$ is viewed as the pure state
vector of the system, from which the eigenstate and the eigen-energy
can be deduced. The single quantum system can be, e.g., one
electron, field, or ensemble of electrons, as long as there exists
coherence. Here, ``single'' is different with that in CM, where the
system is labeled by mass. The wave function has the property of
holistic, and we can say that the system is ``open inside''. Now,
for Schr\"odinger equation, it is easy to get the following
equations
\begin{eqnarray}
i\hbar\dot{|\Psi\rangle}\langle\Psi| &=& \mathcal {H}
|\Psi\rangle\langle\Psi|, \\ \nonumber
-i\hbar|\Psi\rangle\dot{\langle\Psi|} &=&
|\Psi\rangle\langle\Psi|\mathcal {H}^{\dagger},
\end{eqnarray}
where Hamiltonian $\mathcal {H}=\mathcal {H}^{\dagger}$. Let the
density matrix (or statistical matrix)
$\rho=|\Psi\rangle\langle\Psi|$, and
$\dot{\rho}=\dot{|\Psi\rangle}\langle\Psi| +
|\Psi\rangle\dot{\langle\Psi|}$, then
\begin{equation}
  i\hbar\dot{\rho}=[\mathcal {H},\rho],
\end{equation}
which is the Liouville equation.

In reality, it is often impossible to get the wave function, i.e.,
to collect all the information of the system. This can be described
by the well-known $\mathcal{U}=\mathcal{S}+E$ model, where the
universe $\mathcal{U}$ contains one system $\mathcal{S}$ embodied by
one environment $E$. The wave function distributes across
$\mathcal{S}$ and $E$. According to the method of entanglement and
Schmidt decomposition \cite{schmidt}:
\begin{equation}
|\Psi\rangle=\sum_i \lambda_i |s_i\rangle |e_i\rangle,
\end{equation}
then the density matrix
\begin{equation}
\rho=\sum_{i,j} \lambda_i \lambda_j^*|s_i\rangle |e_i\rangle \langle
s_j| \langle e_j|,
\end{equation}
where coherence is the non-diagonal elements. The state of the
system is deduced by tracing out the environment
\begin{equation}
\rho_s=\textrm{tr}_{E}\rho=\sum_i \lambda_i^2|s_i\rangle \langle
s_i|,
\end{equation}
with $\langle s_i|s_j\rangle=\delta_{ij}$, which is often viewed as
the origin of the classical measurement results.

Further, interestingly, the wave function can also be written as
\begin{equation} \label{eq:wfes}
|\Psi\rangle=\sum_k\gamma_k|\psi_k\rangle,
\end{equation}
then the density matrix is
\begin{equation}
\rho=\sum_k p_k|\psi_k\rangle \langle \psi_k|,
\end{equation}
where the set $\{ |\psi_k \rangle \}$ is also wave function, thus,
not necessarily orthogonal, and the parameters $\gamma_k$ are complex, with $|\gamma_k|^2=p_k$.
The state in Eq. (\ref{eq:wfes}) is the ``wave function of ensemble state'' (WFES), the detailed properties are discussed in Appendix \ref{sec:wfesd}.

This decomposition is coarse-grained, i.e., each $|\psi_i \rangle$ can
further be decomposed as the superposition of orthogonal states. In
the ensemble, each party has the property of openness that
there can be (but not necessarily) coherence between any two of
them.

Last, another well-known form is the Wigner function \cite{Wigner}
in the quantum phase-space, which is often written as
\begin{equation}
W(q,p,t)=\frac{1}{\pi \hbar}\int \textrm{d}x e^{-ipx/\hbar}
\psi^*(q-x, t) \psi (q+x, t),
\end{equation}
where the integral is across the whole space. There exists a whole
class of distribution functional, with the Wigner function as the
simplest and straightforward one \cite{Lee95}. Here, in the
phase-space, the system is represented by the coordinate and its
movement is momentum $\bf{p}$. Contrast with classical method, in
Wigner function, there exists another coordinate $x$, which is
integrated out. The physical meaning is that the effects of the
outside world (environment) should be deleted (tracing out), so that
we can confirm the existence of the system. From this point, we can
say that the phase-space approach in QM is also natural since it
captures the essence of openness, the negative value of $W$ is one
expression.

To sum up, in this section we show that the existence of coherence
is central for quantum behaviour, which results in the openness
demonstrated by all the basic equations of QM. We note that here we
do not study the detailed properties and classifications of quantum
coherence. In the next section, we will further show that quantum
coherence and the property of openness can be based on the new
concept of ``scope'', and from which, we will analyze the direct
physical meaning of the wave function and also entanglement.

\section{SCOPE}
\label{sec:SCOPE}

In this section, we study the foundation of QM from the concept of
``scope''. Briefly, scope, or quantum scope, describes the
``structure of motion'', which means the action region of the
motion, and the correlations among states and observable, rather
than the casual dynamics, neither the structure of matter, nor
matter. By comparison, according to CM, the object, labeled by mass
$m$, associates with a certain force, satisfying ${\bf F}=m {\bf
a}$. Here, in QM every motion has a certain scope. The wave function
is indeed the scope of the quantum dynamics. Further, the concept of
scope is not restricted within QM, it is the universal method, and
it has the connections with other methods, such as Relativity.

Below, we present the theorems and properties of scope. Before that,
we first note two points. Firstly, the concept of scope is in
consistent with QM, since it is directly based on QM. Secondly,
scope can give physical meaning to wave function and also
superposition, then complete the basic principle of QM.

\subsection{Theorems} \label{sec:THEOREMS}

Firstly, we present the theorems relating to scope itself, we state
that the principles of superposition and entanglement are universal.

\newtheorem{Theorem}{Theorem}
\begin{Theorem}
Every motion has one scope $\mathcal {S}$.
\end{Theorem}

Here ``motion'' includes the condition when the observed velocity of
one object is zero. Strictly, there is no object at rest. Scope
$\mathcal {S}$ belongs to the motion of a certain object, not the
object itself. The concept of scope describes the ability of motion
in a systematical way.

\begin{Theorem}
Scope $\mathcal {S}$ is composed with state $\psi_i$ and its weight
$\alpha_i$, where $|\langle \psi_i|\psi_{i'} \rangle|\geq 0$, $i=1, 2 ,...n$, $n$ decides the space of $\mathcal
{S}$, labeled as $\mathcal {S}_n\{\alpha_i|\psi_i\rangle\}$.
\end{Theorem}

The concept of ``state'' originates from QM and statistical physics.
$\psi_i$ can be a vector in the Hilbert space, or the point in the
Tangnet space \cite{wang}. Within one scope $\mathcal {S}$, there
can be many physical states $\psi_i$. One motion can only be in
state within its own scope.

The principle of superposition has two parts:

\begin{itemize}
\item
{\em The principle of superposition of scope $\mathcal {S}$}:\\
The scope of motion can be expressed as the superposition of all the
states within as
\begin{equation}
\mathcal {S}_n \equiv \sum^{n}_{i}\alpha_i\psi_i.
\end{equation}
There is a set of POVM ``active operator'' $\mathcal {A}$ which
associates with $\mathcal {S}$ and satisfies

\begin{equation}
\mathcal {A}_i |\mathcal {S}\rangle=\alpha_i|\psi_i\rangle, i=1,
2,...n,
\end{equation}
which means $\mathcal {A}_i$ can act state $|\psi_i\rangle$ out from
scope $\mathcal {S}$. Here, we employ Dirac's ``bra-ket'' symbol
\cite{Dirac}.
\end{itemize}

$\mathcal {A}_i$ satisfies

\begin{equation}
\langle \mathcal{S}|\mathcal {A}_i|\mathcal
{S}\rangle=\alpha_i\langle\mathcal {S}|\psi_i\rangle=|\alpha_i|^2,
\end{equation}
which means $|\alpha_i|^2$ is the eigenvalue of the ``activer''
$\mathcal {A}_i$. This is the alternative of the ``probabilistic
interpretation'' of the coefficients $|\alpha_i|^2$ in QM.

Also
\begin{equation}
\sum_i\langle \mathcal{S}|\mathcal {A}_i|\mathcal
{S}\rangle=\langle\mathcal {S}|\sum_i\mathcal {A}_i|\mathcal
{S}\rangle=1=\sum_i|\alpha_i|^2,
\end{equation}
which means all the states are acted out, that is, the whole scope
$\mathcal {S}$ is realized $\sum_i\langle\mathcal {S}|\mathcal
{A}_i|\mathcal {S}\rangle=1$, thus $\sum_i|\alpha_i|^2=1$. This is
the origin of the ``normalization regulation''. From the
probabilistic view, it means the sum of all the probabilities equals
to 1.

And, $(\mathcal {A}_i+\mathcal
{A}_j)|\mathcal{S}\rangle=\alpha_i|\psi_i\rangle+\alpha_j|\psi_j\rangle=(\mathcal
{A}_j+\mathcal {A}_i)|\mathcal{S}\rangle; \mathcal {A}_i\mathcal
{A}_j|\mathcal{S}\rangle=0, \mathcal {A}_i\mathcal {A}_j=0, i \neq
j; \mathcal {A}_i^2=\mathcal {A}_i.$ Also, there exists the
``anti-active operator'' 1-$\mathcal {A}_i \equiv \mathcal
{\forall}_i$. We do not study the active operator in detail here.

\begin{itemize}
\item
{\em The principle of superposition of state $|\psi\rangle$}:\\
The state of motion can be the coherent superposition state as
\begin{equation}
|\psi\rangle=\sum_j\beta_j|\psi_j\rangle, \sum_j|\beta_j|^2=1,
\end{equation}
where normalization means the norm of a state vector is $1$, which
is the result, yet not the same with the property of scope. Note that
the superposition for mixed, pure, also classical states are all included.
\end{itemize}

\begin{Theorem}
There exists entanglement between two or multi- scopes.
\end{Theorem}

The entangle process $\mathbb{E}$ for two systems
$\Psi_n\{a_i|\psi_i\rangle\}$ and $\Phi_m\{b_j|\phi_j\rangle\}$ can
be written as

\begin{equation}
\mathbb {E} \equiv |\Psi_n\rangle \infty |\Phi_m\rangle \equiv
\sum_{i,j}^{n,m}\mathcal {A}_i^{\Psi}\{a_i|\psi_i\rangle\}\mathcal
{A}_j^{\Phi}\{b_j|\phi_j\rangle\}.
\end{equation}

We note that the analysis can be generalized to multi-systems
directly, here we focus on bi-party system for clarity. $\infty$
represents the entangle process.

There are two parts of the principle of entanglement:

\begin{itemize}
\item
{\em The principle of entanglement of scope $\mathcal {S}$}:\\
The entangled scope of motion can be
\begin{equation}
\mathcal {S}= \sum_{i}^{\textrm{min}\{n,m\}}a_ib_i|\psi_i\rangle
\otimes |\phi_i\rangle /\sqrt{\sum_i a^2_ib^2_i},
\end{equation}
where the sets of relative states \cite{everett}
$\{|\psi_i\rangle\}$ and $\{|\phi_j\rangle\}$ correlate with each
other one-to-one, which can be labeled as the ``1-1 branch'' rule.
The details which $|\psi_i\rangle$ is the relative state to
$|\phi_j\rangle$ is determined by nature, e.g., the interaction.

There also exist the set of active operator $\mathcal {A}_i \equiv
\mathcal {A}_i^{\Psi} \otimes \mathcal {A}_i^{\Phi}$, which
satisfies
\begin{eqnarray}
\mathcal {A}_i |\mathcal {S}\rangle &=&
a_ib_i|\chi_i\rangle/\sqrt{\sum_i a^2_ib^2_i}, \\ \nonumber \langle
\mathcal{S}|\mathcal {A}_i|\mathcal {S}\rangle &=& a_i^2b_i^2/\sum_i
a^2_ib^2_i, \\ \nonumber
\sum_i \langle \mathcal{S}|\mathcal
{A}_i|\mathcal {S}\rangle &=& 1,
\end{eqnarray}
where the branch $|\psi_i\rangle \otimes |\phi_i\rangle \equiv
|\chi_i\rangle$.
\end{itemize}
\begin{itemize}
\item
{\em The principle of entanglement of state $|\psi\rangle$}:\\
The entangled state can be a part of the entangled scope
\begin{equation}
|\Psi\rangle= \sum_{i}a_ib_i|\chi_i\rangle
/\sqrt{\sum_i a^2_ib^2_i}.
\end{equation}
Three kinds of states have to be classified: (1) if $\langle \chi_i|\chi_j\rangle=\delta_{ij}$, the state is truly entangled; if $|\langle \chi_i|\chi_{i'}\rangle| \geq0$, the state is separable, the same as the usual form in Eq. (\ref{eq:separable}); this means the separable state is also the result of the entangle process of scope thus containing quantum information as we will show below; (3) if $i=1$, the state reduces to a simple product state.
\end{itemize}

In reality, the branches within the entangled scope are not all
necessarily realized, thus, the entangled state is the fragment of
the scope as long as satisfying the normalization rule.

In addition, there may be interactions during the entangle process,
however, entanglement describes the information transition or
coherent correlations between different systems, instead of the
energy transition. We will define the degree of entanglement
$\mathcal {E}$ in Sec. \ref{sec:degree}.

Next, relating to QM, particularly the dynamics of the scope and
state, there are two theorems.

\begin{Theorem}
There exist representations, corresponding to different complete
sets formed by the commutative observable, labeled as ``the scope
under representation''.
\end{Theorem}

In QM, the physical reason for representation is that there exist
different complete sets, which is demonstrated by Bohr's
complementarity principle \cite{bohr}. We note that, according to
the consistent history approach \cite{ch}, there exist several
frameworks, which also satisfy the complementarity principle. This
theorem indicates again that the concept of scope is the
systematical, instead of casual, description of the potentiality and
structure of the motion.

\begin{Theorem}
There exist pictures, corresponding to different frames and time
choice.
\end{Theorem}

There are three basic kinds of pictures as follows:

\begin{itemize}
\item {\em The frame is set on the motion itself} : Schr\"odinger
picture.

The scope $\Psi$ (can also be labeled as $\mathcal{S}$) satisfies
the Schr\"odinger equation
\begin{equation}
i\hbar \frac{\partial \Psi}{\partial t}=\mathcal {H}\Psi,
\label{eq:sch}
\end{equation}
where $\mathcal {H}$ is the Hamiltonian.
\end{itemize}

The solution of this equation is the ``structure of the motion''.
Here, ``$t$'' means the change of space rather than the evolution of
time \cite{Zeh01}. In fact, ``$t$'' means the ``gradually evolving
of the scope''. The dynamical variables do not change with time.

We mention that in quantum cosmology, Eq. (\ref{eq:sch})
becomes $\mathcal {H}\Psi=0$, where the Hamiltonian includes the
gravitational field plus all matter sources in the universe, the
external timing parameter $t$ vanishes naturally \cite{ch}.
\begin{itemize}
\item {\em The frame is set external}: Heisenberg picture.

The dynamical variable $\hat{A}(t)$ satisfies
\begin{equation}
i\hbar\dot{\hat{A}}(t)=-[\mathcal {H},\hat{A}(t)],
\end{equation}
which means the change of dynamical variable in different states
within the fixed scope.
\end{itemize}
\begin{itemize}
\item {\em The frame is set in-between}: Dirac (interaction) picture \cite{Dirac}.

$\Psi_I(t)=e^{i\mathcal {H}_0t/\hbar}\Psi_S(t)$,
$\hat{A}_I(t)=e^{i\mathcal {H}_0t/\hbar}\hat{A}_S e^{-i\mathcal {H}_0t/\hbar}$,
satisfying
\begin{eqnarray}
i\hbar \dot{\Psi}_I(t)&=&\mathcal{H}_I \Psi_I(t),\\ \nonumber
i\hbar\dot{\hat{A}}_I(t)&=&-[\mathcal {H},\hat{A}_I(t)],
\end{eqnarray}
where $\mathcal {H}=\mathcal {H}_0+\mathcal {H'}$.
\end{itemize}

The existence of pictures and the problem of the external parameter
$t$ reveal that scope mostly describes the logic structure of the
states, from which the commutation relations of observable do not
change with time.

Last, we show another kind of property of scope. There exist
relations between the concept of scope and Relativity, which
describes the coherent properties of the space-time metric of the
object. Here, we only present the mostly crucial conjecture as
below.

\begin{Theorem}
The scope $\mathcal {S}$ can be influenced by mass $\mathcal {M}$.
\end{Theorem}

This theorem sets the connection between the motion of a certain
object and the object itself. It means that objects with different
masses can not have the same scope. Every scope is unique. Also, the
shape of scope is not regularly euclidian. In general, it is
topologically non-euclidian. Please see the geometrical properties
$1-3$ of scope below.

\subsection{Properties} \label{sec:PROPERTIES}

In this subsection, we study some properties and consequences
deduced from the theorems above of scope.

\newtheorem{Property}{Property}
\begin{Property}
Scope $\mathcal {S}$ has a certain geometrical shape.
\end{Property}

The number of state $n$ can not be zero. The states within the scope
can form certain shape. For instance, if there is one single state,
$n$=1, the corresponding shape of scope is just a point (in
geometry). If $n$=2, the shape is one segment line, straight or
carved, with a certain length. If $n$=3, the shape is one triangle
face, with a certain area. If $n$=4, the shape is one tetrahedron.
If $n\rightarrow \infty$, the shape is one ball. We should state
that the shape is in the three-dimensional space, euclidian or
non-euclidian, different from the shape in the high-dimensional
space, such as the Hilbert space. The shape not only has the
``shell'' with boundary and face, also it has inner structure.
Particularly, when $n\rightarrow \infty$, if we neglect its
structure, the ball can be viewed as one point. This is the
alternative of the ``classical limit'' of the orthodox QM by taking
$\hbar \rightarrow 0$.

We note that the geometric properties of quantum states are under
research, see Ref. \cite{geo} for instance.

\begin{Property}
Scope $\mathcal {S}$ has a certain magnitude.
\end{Property}

The magnitude is defined as the length (or the area, the volume) of
its shape. To calculate, we have to specialize the ``distance''
between different states. For instance, we can use the energy $E$ to
define the distance $\mathcal {D}$, such as $\mathcal
{D}(\psi_i,\psi_j)=|E_i-E_j|.$

The magnitudes of $\mathcal {S}$ of different movements should be
different, since they have physical meanings. The bigger the
magnitude, the better the ability of the motion, thus, the more
``freedom'' in the scope $\mathcal {S}$. The magnitude is not
necessarily equal to one, which is different with the normalization
condition.

\begin{Property}
One particular scope $\mathcal {S}$ has its own space-time matric.
\end{Property}

This is the direct result of Theorem $6$. When the effect of mass is
not obvious, we can set the space-time matric of different scopes
the same with each other, which is another kind of classical limit,
or the low-energy limit.

\begin{Property}
The Scope $\mathcal {S}$ has the properties of ``reality''
(actuality) and ``propensity''.
\end{Property}

When $\mathcal {A}_i$ acts, the certain state $|\psi_i\rangle$ is
realized, from potential to real, and the scope $\mathcal {S}$
``gradually evolves to its whole form''. There is the transition
between reality and propensity. Actually, the two aspects of scope
were studied a long time ago by Aristotle in the ancient Greece
\cite{Aristotle}. The method was mostly ignored and was claimed as
``metaphysical'' useless for physics. Yet, the idea of propensity
has gained quite a lot of attentions in recent years
\cite{Aerts,Ronde,heisenberg}.

The concept of $\mathcal {S}$ manifests that QM is the systematical
description of the structure of movement, which is totally new
compared with the classical dynamical method, from which the ``local
realism'' \cite{epr} originates. Thus, we view the critics based on
local realism as the misunderstanding of the essence of QM. Here, we
do not discuss the problems of local realism and nonlocality in
detail.

Also, there is no ``collapse'' \cite{neumann}. Collapse means that
when there is measurement, the state collapses from superposed state
to the eigenstate. Or, some others view the collapse happens in our
knowledge \cite{espagnat}. In fact, with the process of decoherence
\cite{zurek03}, when the initial state is superposed, it decoheres
instead of collapse when there is measurement or environment.

\begin{Property}
Measurement is a process of entanglement.
\end{Property}

This is a simple result of the fact that the apparatus also has one
scope with the states relative to the observed system. The
measurement process is the disturbance of system to apparatus (or
environment) \cite{zurek03}. The measurement problem in QM
\cite{Wallace}, as well as the observe effect in Relativity, can be
well explained with the method of entanglement, and the change of
our knowledge is another story \cite{espagnat}.

\begin{Property}
There are mainly two kinds of states: time-quasistatic (Tq) state
and ensemble-isotactic (Ei) state.
\end{Property}

At present, there is no unified classification of quantum states.
Briefly, there are many kinds due to different criterions, such as
separable/entangled, pure/mixed, local/non-local states, etc. The
non-unification just indicates that the underlying physical reason
for the classification is not clear.

We give a new classification as follows:

\begin{itemize}
\item onefold state
\begin{itemize}
\item eigenstate
\item superposed state
\begin{itemize}
\item time-quasistatic (Tq) superposed state
\item ensemble-isotactic (Ei) superposed state
\end{itemize}
\end{itemize}
\item multifold state
\begin{itemize}
\item factorizable state: product, separable, etc
\item entangled state
\begin{itemize}
\item time-quasistatic (Tq) entangled state
\item ensemble-isotactic (Ei) entangled state
\end{itemize}
\end{itemize}
\end{itemize}

Firstly, state is differentiated according to whether it is a
multifold (many-body) state. For a single object, its state is
onefold state. There can be superposed state according to the
principle of superposition. For multifold system which contains
several inner parts, the state is multifold state, and there can be
entangled state. The separable state can be viewed as a kind of
factorizable state, which we will study in the next section. The
distinction of time-quasistatic (Tq) state and ensemble-isotactic
(Ei) state bases on the theorem of ergodicity of statistical physics
and the principle of identity in QM \cite{Fano}.

For example, the double-slit state of the electron (or other
particles) is the Ei superposed state, which characterizes the
electron ensemble. In this state, for one particular electron, it
can only pass through one slit at one time, thus, the state of a
single electron is not superposed. Another example, as we know, we
can use laser to control a certain two-level atom, and drive it to
the so-called Rabi oscillation \cite{Rabi}. This is the Tq state,
the population transfers from one state to another periodically. The
famous Bell basis are the Ei entangled state, such as the light
source from type-II SPDC \cite{Kwiat} in the teleportation
experiment, describing the ensemble of photons, the state of one
photon is not entangled or superposed. An excellent example of Tq
entangled state is the electrons forming the chemical bonding in the
molecules.

The difference between (Ei) state and (Tq) state is the ``inner''
dynamics of the state itself, thus, this kind of definition is
obviously not mathematical. However, in our study below, we do not
care about this difference without loss of generality.

\section{ENTANGLEMENT}
\label{sec:ENTANGLEMENT}

\subsection{General remarks} \label{sec:REMARKS}

\begin{figure}
\includegraphics[scale=0.3]{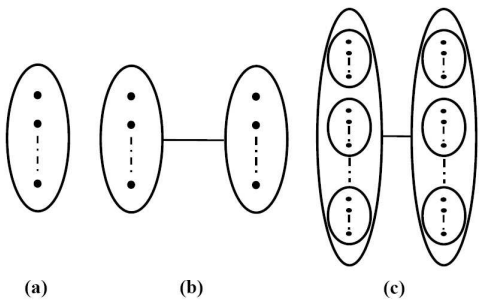}
\caption{The diagrams for the single quantum state (a), product
state (b), and ensemble-product state (c). The circle stands for the
scope, and the points within stand for the orthogonal eigenstates
(the total number is assumed to be $d$, thus, qu$d$it). The
correlation between scopes or states is depicted as straight line
(string).} \label{fig.product}
\end{figure}

QM often deals with the properties of many-body (instead of single
or infinite) system. To characterize the quantum features, at
present, there exist three widely studied methods: nonlocality
\cite{bell87}, nonclassicality \cite{glauber}, and entanglement
\cite{horodecki}. Traditionally, nonclassicality was introduced in
the quantum phase-space in the context of Quantum Optics. The state
of light, e.g., squeezed light, contains nonclassicality compared
with the coherent light, which sets the boarder between quantum and
classical states. We note that the continuous variable (CV) system
in phase-space can be translated into the Fock space, from which the
entanglement can be defined. Interestingly, after the seminal work
of EPR \cite{epr} and Bell \cite{bell87}, the nonlocality of
entangled state was demonstrated with nonclassical light
\cite{aspect}, where the three methods are all involved. In this
work, we do not intend to analyze the relations of them; instead, we
focus on entanglement relating to superposition and the concept of
scope. Part of our study on entanglement has been presented in Ref.
\cite{wang}. We will study the physical role of entanglement in
various quantum states, and the degree of entanglement, also we
compare with different entanglement measure in Appendix
\ref{sec:MEASURE}.

Entanglement is widely studied in QIQC, and it is believed to be the
key for information processing and computing. However, entanglement
is not indispensable. In quantum cryptography, the BB$84$ does not
rely on entanglement \cite{bb84}; instead, particularly the B$92$
protocol shows that it is the nonorthogonality (also no-cloning)
that ensures the security of the key \cite{b92}. The recently
studied continuous variable (CV)-quantum key distribution (QKD)
\cite{Grosshans} and decoy-state QKD \cite{Lo}, also do not rely on
entanglement, which indeed are the improvement of the BB$84$
protocol. This indicates that the quantum nature of information acts
even when there is no entanglement, i.e., the quantum information
does not necessarily depends on entanglement. In quantum computing,
the well-known QDC$1$ model is shown not directly rely on
entanglement \cite{Knill}; instead, the quantum discord plays the
central role \cite{Datta}. In another line of research, there are
attempts to build QKD on the foundation of nonlocality and nonlocal
box, where entanglement does not play the central role
\cite{popescu,Barrett}.

The examples above indicate that entanglement is not such quantity
viewed from the information-theoretic framework (ITF), which is the
theoretic paradigm or the world view correlating with the research
of QIQC to re-consider quantum theory and its foundation from the
information point, where information is believed as physical and
fundamental \cite{Bub,Wheeler1,Wheeler2,Landauer}. On the contrary,
especially from the concept of scope, entanglement is the direct
generalization of superposition, not necessarily related to
information (entropy), physically. The original CV-entangled state
of EPR \cite{epr} and (discrete variable) DV-entangled state of spin
of Bohm \cite{Bohm} do not relate to information. The analysis of
Schr\"odinger just reveals that information can be effected by
entanglement \cite{Schrodinger}. Also, entanglement does not only
play roles in QIQC. For instance, in quantum chemistry, entanglement
plays roles in the formation of molecules and chemical reactions.
Further, the so-called ``quantum information'' is quite misleading,
as the Zeilinger's principle \cite{Zeilinger99} states that there is
no quantum information, there is only information processing with
quantum sources. That is to say, information can be effected by the
quantum features, including nonorthogonality (superposition),
entanglement, also nonlocality and nonclassicality.

\begin{figure}
\includegraphics[scale=0.35]{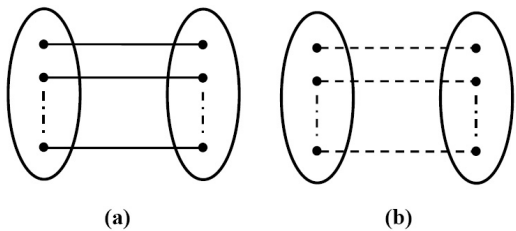}
\caption{The diagrams for the entangled qudit (a), and decohered
qudit (b). The straight lines (string) stand for the correlation,
the dashed lines stand for decohered correlation.}
\label{fig.entangled}
\end{figure}

\begin{figure}
\includegraphics[scale=0.3]{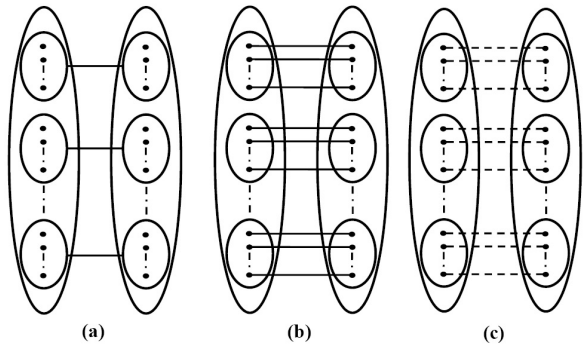}
\caption{The diagrams for the separable state (a),
ensemble-entangled qudit (b), and ensemble-decohered qudit (c).}
\label{fig.ensemble}
\end{figure}

\subsection{Quantum states} \label{sec:STATES}

In this section, we will systematically study the quantum
correlations of the bi-party system, the results for the multi-party
system will be presented in other places. Also, without loss of
generality, we assume here that the branches within the entangled
scope are all realized, i.e. mathematically, the entangled state is
the same with the entangled scope. Relating to the analysis in Sec.
\ref{sec:ENSEMBLE}, the entangled state can be realized both for
single quantum system and the ensemble of quantum system. The
quantum states we analyze include:

\begin{itemize}
\item product state; entangled qudit; decohered qudit;
\item ensemble-product state; separable state; ensemble-entangled qudit; ensemble-decohered
qudit.
\end{itemize}

For clarity of the physical picture, we develop a kind of diagram to
represent the scope. As shown in Fig. \ref{fig.product}(a), the
circle stands for the scope, and the points within stand for the
orthogonal eigenstates (with number $d$). The correlation between
scopes or states is depicted as straight line (string), the dashed
line stands for decohered string. The states we consider are shown
in Fig. [\ref{fig.product}-\ref{fig.ensemble}].

{\em Product state.} As well-known, product state contains no
correlation. Here, we can describe three kinds of product states.
Suppose the two sub-systems are $A$ and $B$, with orthogonal
eigenstate $\{ A_i \}$ and $\{ B_i \}$, respectively.

(a). The state $|A_i\rangle \otimes |B_i\rangle$ is product state.

(b). The more complex product form is
\begin{equation}
\rho_P=\rho_A \otimes \rho_B= |\psi_A \rangle \langle \psi_A|
\otimes |\psi_B \rangle \langle \psi_B|, \label{eq:product}
\end{equation}
where $|\psi_A \rangle=\sum_i a_i | A_i \rangle$, $|\psi_B
\rangle=\sum_i b_i | B_i \rangle$. We note that the states for $A$
and $B$ are not ensemble. This product state is shown in Fig.
\ref{fig.product}(b).

(c). The third kind is the ensemble-product state, shown in Fig.
\ref{fig.product}(c). The wave functions of the systems $A$ and $B$
can be written as
\begin{eqnarray}
\nonumber |\psi_A \rangle &=& \sum_{\xi} \gamma_{\xi} |
\psi_{\xi}^A \rangle, \\ |\psi_B \rangle &=& \sum_{\xi}
\gamma_{\xi} | \psi_{\xi}^B \rangle,\label{eq:cg1}
\end{eqnarray}
which is the coarse-grained decomposition, since there exist the
sub- wave functions $|\psi_{\xi}^A \rangle=\sum_i a^{\xi}_i | A_i
\rangle$, $|\psi_{\xi}^B \rangle=\sum_i b^{\xi}_i | B_i \rangle$,
and generally $\langle \psi_{\xi}^{A(B)}|\psi_{\xi '}^{A(B)} \rangle
\neq 0$. With $p_{\xi}=|\gamma_{\xi}|^2$, the density matrix can be easily written as
\begin{eqnarray}
\nonumber \rho_A &=& \sum_{\xi} p_{\xi} | \psi_{\xi}^A
\rangle\langle
\psi_{\xi}^A| = \sum_{\xi} p_{\xi} \rho_{\xi}^A, \\
\rho_B &=& \sum_{\xi} p_{\xi} | \psi_{\xi}^B \rangle\langle
\psi_{\xi}^B| = \sum_{\xi} p_{\xi} \rho_{\xi}^B. \label{eq:cg2}
\end{eqnarray}
The ensemble-product state is
\begin{equation}
\rho_{eP}=\rho_A \otimes \rho_B= ( \sum_{\xi} p_{\xi} \rho_{\xi}^A)
\otimes (\sum_{\xi} p_{\xi} \rho_{\xi}^B). \label{eq:cg3}
\end{equation}
The state $\rho_{eP}$ contains quantum information, although there
is no entanglement. There is coherence for the $\rho_{\xi}^A$ and
$\rho_{\xi}^B$, that is, there can be decoherence process. The
result of the decoherence is that the local state of $A$ and $B$
become $\rho_{\xi}^A=\sum_i (a^{\xi}_i)^2 | A_i \rangle \langle
A_i|$, and $\rho_{\xi}^B=\sum_i (b^{\xi}_i)^2 | B_i \rangle \langle
B_i|$. Yet, the global state $\rho_{ep}$ remains as product.

{\em Entangled qudit.} The general form of the entangled qudit is
written as
\begin{equation}
|\Psi_E\rangle =\frac{\sum_i a_ib_i|A_i\rangle \otimes | B_i
\rangle}{\sqrt{\sum_i a_i^2b_i^2}},
\end{equation}
and the density matrix is expressed as
\begin{equation}
\rho_E =|\Psi_E\rangle \langle \Psi_E |= \frac{\sum_{i,j} \eta_i
\eta_j|A_i\rangle \langle A_j|\otimes | B_i \rangle \langle
B_j|}{\sum_i \eta_i^2}, \label{eq:entan}
\end{equation}
with $\eta_i=a_ib_i$.

The state $\rho_E$, shown in Fig. \ref{fig.entangled}(a), can be
viewed as a part of the product state $\rho_P$ in case (b) above, by
selecting out the branches according to the permutation symmetry.
The coherence of the state $\rho_E$ is represented by the
non-diagonal elements. When decoherence occurs by some disturbance,
the state $\rho_E$ becomes diagonal, which is said to be classical.
The decohered (classical) qudit is
\begin{eqnarray}
\nonumber \rho_D &=& \sum_i \eta^2_i | A_i \rangle\langle A_i|
\otimes | B_i \rangle\langle B_i| /\sum_i \eta^2_i \\ &=& \sum_i
a_i^2 b_i^2 \rho_i^A \otimes \rho^B_i/\sum_i a^2_i b^2_i.
\label{eq:decoher}
\end{eqnarray}
Let $\rho_A=\sum_i a_i^2 \rho_i^A$, $\rho_B=\sum_i b_i^2 \rho_i^B$,
and introduce the parameter $p_i=\eta^2_i/\sum_i \eta^2_i$, then
\begin{equation}
\rho_D = \sum_i p_i \rho_i^A \otimes \rho_i^B,
\end{equation}
which is separable. That is, for the single decohered qudit, it is
classical as well as separable. This state is shown in Fig.
\ref{fig.entangled}(b).

{\em Separable state.} The well-known separable state
\cite{Werner89} in Fig. \ref{fig.ensemble}(a) is defined as
\begin{equation}
\rho_S = \sum_{\xi} p_{\xi} \rho_{\xi}^A \otimes \rho_{\xi}^B,
\label{eq:separable}
\end{equation}
with the normalization relation $\sum_{\xi} p_{\xi}=1$.

We show that the separable state $\rho_S$ is the analogy of
entangled state $\rho_E$, except that the former one employs the
coarse-grained decomposition, while the latter one employs the
fine-grained decomposition, i.e., the eigenstates are orthogonal.
The decomposition is the same with those in equations (\ref{eq:cg1})
and (\ref{eq:cg2}). Then equation (\ref{eq:separable}) can be
written as
\begin{equation}
\rho_S = \sum_{\xi} p_{\xi} (\sum_{k,l}a_k^{\xi}a_l^{\xi *}
|A_k\rangle \langle A_l|)\otimes(\sum_{m,n}b_m^{\xi}b_n^{\xi *}
|B_m\rangle \langle B_n|), \label{eq:separable2}
\end{equation}
which is the analogy of equation (\ref{eq:entan}).

The ensemble (sets) $\{|\psi_{\xi}^A \rangle \}$ and
$\{|\psi_{\xi}^B \rangle\}$ are not classical, while the ensemble
$\{|A_i \rangle \}$ and $\{|B_i \rangle\}$ are classical. This
reveals that there exists quantum information in the separable
state. Then, when the separable state goes to the classical state?
Consider the case when only one eigenstate each is acted out, that
is $d=1$, and also let $\langle \psi_{\xi}^{A(B)}
|\psi_{\xi'}^{A(B)} \rangle =0$. Then the density matrix $\rho_S$ in
equation (\ref{eq:separable2}) becomes the same with the classical
state $\rho_D$ in equation (\ref{eq:decoher}). We can label this as
the ``classical reduction procedure''.

Let us give a simple example of the $2\otimes 2$ system. It is
reasonable
to set $|\psi_1^{A(B)} \rangle=\left( \begin{array}{c}1\\
0\end{array} \right)$, and $|\psi_2^{A(B)} \rangle=\left( \begin{array}{c}0\\
1\end{array} \right)$, then the separable state is
\begin{equation}
\rho_S = \left( \begin{array}{cccc}p_1 & 0&0& 0\\ 0& 0& 0& 0\\ 0 &0& 0& 0\\
0& 0& 0& p_2\end{array} \right),
\end{equation}
which is also classical.

Generally, when $d \neq 1$, the separable form of the density matrix
does not rely on the local coherence within $\rho_{\xi}^A$ and
$\rho_{\xi}^B$, which just indicates that there can be quantum
information in the separable state.

{\em Ensemble-entangled qudit.} The ensemble of entangled state is
often studied in QIQC, such as the ensemble of Bell's state. For
single entangled qudit, it can be written in the Schmidt basis as
\begin{equation}
|\psi_{\xi}\rangle= \sum_i \lambda^{\xi}_i|A_i\rangle |B_i\rangle,
\end{equation}
with $\sum_i (\lambda^{\xi}_i)^2=1$. The density matrix of this
ensemble can be expressed as
\begin{eqnarray}
\rho_{eE}&=&\sum_{\xi} p_{\xi} |\psi_{\xi}\rangle \langle
\psi_{\xi}| \\ \nonumber &=& \sum_{\xi} p_{\xi} \sum_i
\lambda^{\xi}_i|A_i\rangle |B_i\rangle \sum_j \lambda^{\xi *}_j
\langle A_j | \langle B_j | \\ \nonumber &=& \sum_{\xi} p_{\xi}
\sum_{i,j} \lambda^{\xi}_i \lambda^{\xi *}_j |A_i\rangle \langle A_j
|\otimes|B_i\rangle \langle B_j | \\ \nonumber &\neq& \rho_S.
\end{eqnarray}

The ensemble-entangled qudit is shown in Fig. \ref{fig.ensemble}(b).
It is easy to see that when applies the classical reduction
procedure, $d=1$, and $\langle \psi_{\xi}^{A(B)} |\psi_{\xi'}^{A(B)}
\rangle =0$, the state $\rho_{eE}$ reduces to classical and also
separable.

When decoherence occurs, each entangled qudit becomes classically
correlated, and the ensemble-decohered qudit is
\begin{equation}
\rho_{eD}= \sum_{\xi} p_{\xi} \sum_i (\lambda^{\xi}_i)^2 |A_i\rangle
\langle A_i |\otimes|B_i\rangle \langle B_i |,
\end{equation}
which is classical but not separable, shown in Fig.
\ref{fig.ensemble}(c).

Further, to make the distinctions of the quantum states clear, we
can classify four kinds of correlations: entanglement, decohered
classicality, nonorthogonality, and coarse-grained classicality. The
quantum characters in the states we studied are different. In the
single entangled qudit, the entanglement is the global shared
coherence, there is also the classical correlation. For the
decohered qudit, the shared coherence disappears, thus there is only
classicality. The ensemble-entangled qudit, which relies on the
coarse-grained decomposition, contains another kind of shared
coherence due to nonorthogonal local states, namely, $\langle
\psi_{\xi}^{A(B)}|\psi_{\xi '}^{A(B)}\rangle \neq 0$, we name it as
nonorthogonality. Correspondingly, there exists another kind of
classicality, via the parameter $p_{\xi}$, instead of the
correlations of the eigenstates between $A$ and $B$. We name this
kind of correlation as coarse-grained classicality, and the former
one as decohered classicality. From this analysis, we can briefly
present the main characters of quantum state in TABLE I.

\begin{widetext}
\begin{table*}[!htbp]
\centering \label{tab:1} \tabcolsep 2pt \caption{The main characters
of the quantum states. The four kinds of correlations are shown on
the left, the six kinds of quantum states are shown on the top.
$\surd$ ($\times$) means the corresponding state has (does not have)
the corresponding correlation.} \vspace{1.5pt}
\begin{center}
\begin{tabular}[t]{|c|c|c|c|c|c|c|}\hline
%\begin{tabular*}{\temptablewidth}{@{\extracolsep{\fill}}|c|c|c|c|c|c|}
& qudit  & decohered qudit  & ensemble qudit  & ensemble-decohered qudit & separable  & ensemble-product \\
\hline
entanglement & $\surd$  & $\times$ & $\surd$ & $\times$ & $\times$ & $\times$ \\
\hline
decohered classicality &          $\surd$  & $\surd$  & $\surd$ & $\surd$  & $\times$ & $\times$ \\
\hline
nonorthogonality &          $\times$ & $\times$ & $\surd$ & $\surd$  & $\surd$  & $\surd$ \\
\hline coarse-grained classicality &          $\times$ & $\times$ &
$\surd$ & $\surd$  & $\surd$  & $\times$ \\ \hline
\end{tabular}
%{\rule{\temptablewidth}{1.2pt}}
\end{center}
\end{table*}
\end{widetext}

\subsection{Degree of superposition and entanglement}\label{sec:degree}

In this section, we study superposition and entanglement
quantitatively. Before our analysis, we note that the superposition
and entanglement are both due to the existence of coherence, local
and global (shared), respectively. Entanglement is not the quantum
information; instead, the classical information (correlation) can be
effected by the entanglement (shared coherence), thus, the so-called
``quantum information''. We can say that entanglement and quantum
information (entropy) are the two related kinds of characters of the
entangled state. Below, we present the definitions of the degree of
superposition, entanglement, and nonorthogonality, which all relate
to the quantum coherence.\\

\newtheorem{Definition}{Definition}
\begin{Definition}
Degree of superposition $\varepsilon$. \\

For a superposed state
vector $|\Psi\rangle=\sum_i^n a_i|\psi_i\rangle$, where $\sum_i^n
|a_i|^2=1$, $\langle\psi_i|\psi_j\rangle=0$, the degree of
superposition
\begin{equation}
\varepsilon \equiv \sum_{i<j}^n |a_i||a_j|.
\end{equation}
\label{def:1}
\end{Definition}

Notice that, by introducing the $l_2$-norm as
$||a||_{l_2}$$=\sum_i^n |a_i|^2$, then the $l_1$-norm is
$||a||_{l_1}$$=\sum_i^n |a_i|$, thus, $\varepsilon =
\frac{1}{2}{(||a||_{l_1}^2-||a||_{l_2})}$. This indicates that the
$l_1$-norm has physical meaning, i.e., describing the degree of
superposition.

Let us show some examples. When $n=1$, $\varepsilon=0$. $n=2$,
$|\Psi\rangle=a_1|\psi_1\rangle+a_2|\psi_2\rangle$,
$\varepsilon=|a_1a_2|$, and Max($\varepsilon$)$=1/2$ when
$|a_1|=|a_2|=1/\sqrt{2}$. $n=3$,
$|\Psi\rangle=a_1|\psi_1\rangle+a_2|\psi_2\rangle+a_3|\psi_3\rangle$,
$\varepsilon=|a_1a_2|+|a_2a_3|+|a_1a_3|$, and Max($\varepsilon$)$=1$
when $|a_1|=|a_2|=|a_3|=1/\sqrt{3}$. Then generally, for
$n=4,5,...$, Max($\varepsilon$)$=3/2, 2, ...$ etc. That is, for $n$,
the maximal degree of superposition $\varepsilon$ is
$\frac{n-1}{2}$, and the minimum approaches to {\em zero}. If there
are more eigenstates in a superposed state, then the degree of
superposition it can get is bigger. It means, physically, that
although a state vector in the Hilbert space has to be normalized,
the degree of superposition is not a relative quantity, it
characterizes the amount of coherence.

\begin{Definition}
Degree of entanglement $\mathcal{E}$.\\

For a general entangled state with $m$ subsystems $n \otimes n
\otimes ...\otimes n$ with each $n$-dimensional:
$|\Psi\rangle=\sum_i^n c_i|\psi_{1i}\psi_{2i}...\psi_{mi}\rangle$,
the degree of entanglement
\begin{equation}
\mathcal{E} \equiv \sum_{i<j}^n |c_i||c_j|.
\end{equation}
\label{def:2}
\end{Definition}

The same with the degree of superposition, we can get $\mathcal{E} =
\frac{1}{2}{(||c||_{l_1}^2-||c||_{l_2})}$.

Our definition $2$ is similar with negativity (see Appendix
\ref{sec:MEASURE}), thus, we do not present the proof of the
entanglement monotone \cite{horodecki}, which states that the
entanglement cannot increase under the local quantum operation and
classical communication (LOCC). According to our understanding based
on scope and coherence, the LOCC constraint is the property of
quantum information (entropy) instead of entanglement (globally
shared coherence). Yet, it is easy to check that entanglement
obviously satisfies the LOCC constraint since the quantum coherence
cannot be created via classical way. The definition $2$ defines
entanglement in a positive way. As a result, the LOCC constraint is
not sufficient to define entanglement, and not all non-separable
states are entangled (Sec. \ref{sec:STATES}).

Next we study some basic applications. We assume the coefficients
are all positive, without loss of generality.

For $2\otimes 2$ system, if
\begin{eqnarray}
|\Psi_A\rangle &=& a_1|\psi_1\rangle+a_2|\psi_2\rangle,
\varepsilon_A=a_1a_2, \\ \nonumber |\Psi_B\rangle &=&
b_1|\psi_1\rangle+b_2|\psi_2\rangle, \varepsilon_B=b_1b_2. \\
\nonumber
\end{eqnarray}
Label $|\psi_1\rangle=|1\rangle$, $|\psi_2\rangle=|2\rangle$. There
are two kinds of entangled state according to the nature of the
branches (strings). If $|1(2)\rangle$ of $A$ correlates with
$|1(2)\rangle$ of $B$, it is named as ``direct'' ($|\Psi_d\rangle$);
if $|1(2)\rangle$ correlates with $|2(1)\rangle$, it is named as
``cross'' ($|\Psi_c\rangle$):
\begin{eqnarray}
|\Psi_d\rangle &=& \frac{a_1b_1|1 1\rangle+a_2b_2|2
2\rangle}{\sqrt{a_1^2b_1^2+a_2^2b_2^2}}, \\
\nonumber |\Psi_c\rangle &=&
\frac{a_1b_2|1 2\rangle+a_2b_1|2 1\rangle}{\sqrt{a_1^2b_2^2+a_2^2b_1^2}},\\
\nonumber
\end{eqnarray}
and the degree of entanglement
\begin{eqnarray}
\mathcal{E}_d &=& \frac{a_1a_2b_1b_2}{a_1^2b_1^2+a_2^2b_2^2}, \\
\nonumber \mathcal{E}_c &=&
\frac{a_1a_2b_1b_2}{a_1^2b_2^2+a_2^2b_1^2},\\
\nonumber
\end{eqnarray}
generally, $\mathcal{E}_d \neq \mathcal{E}_c$.

After some algebra, we can get
\begin{equation}
\frac{\varepsilon_A\varepsilon_B}{\mathcal{E}_d}+\frac{\varepsilon_A\varepsilon_B}{\mathcal{E}_c}=1.
\end{equation}
Introduce the reduced degree of entanglement
\begin{equation}
\mathcal{E}^{\dagger}=\frac{\mathcal{E}_d\mathcal{E}_c}{\mathcal{E}_d+\mathcal{E}_c},
\end{equation}
then we can get the fundamental relation
\begin{equation}
\mathcal{E}^{\dagger}=\varepsilon_A\varepsilon_B.
\end{equation}
When $a_1=a_2=b_1=b_2=\sqrt{2}/2$,
$\varepsilon_A=\varepsilon_B=1/2$,
$|\Psi_d\rangle=\frac{\sqrt{2}}{2}{(|1 1\rangle+|2 2\rangle)}$,
$|\Psi_c\rangle=\frac{\sqrt{2}}{2}{(|1 2\rangle+|2 1\rangle)}$,
$\mathcal{E}_d=\mathcal{E}_c=1/2$, and $\mathcal{E}^{\dagger}=1/4$.
This is the maximal condition for the qubit.

In addition, for the reduced degree of entanglement of the qubit
system, we will further study the physical properties by comparing
to the classical two-body problem in Appendix \ref{sec:REDUCED}.

For $2\otimes 2 \otimes 2$ system, if
\begin{eqnarray}
|\Psi_A\rangle &=& a_1|\psi_1\rangle+a_2|\psi_2\rangle,
\varepsilon_A=a_1a_2, \\ \nonumber |\Psi_B\rangle &=&
b_1|\psi_1\rangle+b_2|\psi_2\rangle, \varepsilon_B=b_1b_2, \\
\nonumber |\Psi_C\rangle &=&
c_1|\psi_1\rangle+c_2|\psi_2\rangle, \varepsilon_C=c_1c_2. \\
\nonumber
\end{eqnarray}
There are four kinds of entangled state, which belongs to the
Greenberger-Horne-Zeilinger (GHZ) state \cite{ghz}:
\begin{eqnarray}
|\Psi_{GHZ}^1\rangle &=& \frac{a_1b_1c_1|1 1 1\rangle+a_2b_2c_2|2
2 2\rangle}{\sqrt{a_1^2b_1^2c_1^2+a_2^2b_2^2c_2^2}}, \\
\nonumber |\Psi_{GHZ}^2\rangle &=& \frac{a_1b_1c_2|1 1
2\rangle+a_2b_2c_1|2
2 1\rangle}{\sqrt{a_1^2b_1^2c_2^2+a_2^2b_2^2c_1^2}}, \\
\nonumber |\Psi_{GHZ}^3\rangle &=& \frac{a_2b_1c_1|2 1
1\rangle+a_1b_2c_2|1
2 2\rangle}{\sqrt{a_2^2b_1^2c_1^2+a_1^2b_2^2c_2^2}}, \\
\nonumber |\Psi_{GHZ}^4\rangle &=& \frac{a_1b_2c_1|1 2
1\rangle+a_2b_1c_2|2
1 2\rangle}{\sqrt{a_1^2b_2^2c_1^2+a_2^2b_1^2c_2^2}}, \\
\nonumber
\end{eqnarray}
and it is easy to write the four degree of entanglement (we do not
present here), after some algebra, we get
\begin{equation}
\frac{\varepsilon_A\varepsilon_B\varepsilon_C}{\mathcal{E}_1}+\frac{\varepsilon_A\varepsilon_B\varepsilon_C}{\mathcal{E}_2}+
\frac{\varepsilon_A\varepsilon_B\varepsilon_C}{\mathcal{E}_3}+\frac{\varepsilon_A\varepsilon_B\varepsilon_C}{\mathcal{E}_4}=1.
\end{equation}
Also, introduce the reduced degree of entanglement
\begin{equation}
\mathcal{E}^{\dagger}=\frac{\mathcal{E}_1\mathcal{E}_2\mathcal{E}_3\mathcal{E}_4}
{\mathcal{E}_1\mathcal{E}_2\mathcal{E}_3+\mathcal{E}_1\mathcal{E}_2\mathcal{E}_4+\mathcal{E}_1\mathcal{E}_3\mathcal{E}_4+
\mathcal{E}_2\mathcal{E}_3\mathcal{E}_4},
\end{equation}
then
\begin{equation}
\mathcal{E}^{\dagger}=\varepsilon_A\varepsilon_B\varepsilon_C.
\end{equation}

For the maximal condition, when all the coefficients equal to
$\sqrt{3}/3$, $\varepsilon_A=\varepsilon_B=\varepsilon_C=1/2$,
$\mathcal{E}_1=\mathcal{E}_2=\mathcal{E}_3=\mathcal{E}_4=1/2$, and
$\mathcal{E}^{\dagger}=1/8$.

Generally, for $2~ \otimes ^n$ system, the maximal reduced
entanglement is $\mathcal{E}^{\dagger}=1/2^n$. This means when we
add the numbers of systems to be entangled, the reduced entanglement
becomes smaller and smaller, that is, the entanglement of the whole
system becomes more fragile, when one subsystem is disturbed, the
total entanglement vanishes.

For $3 \otimes 3$ system, if
\begin{eqnarray}
|\Psi_A\rangle &=&
a_1|\psi_1\rangle+a_2|\psi_2\rangle+a_3|\psi_3\rangle,
\\ \nonumber |\Psi_B\rangle &=&
b_1|\psi_1\rangle+b_2|\psi_2\rangle+b_3|\psi_3\rangle, \\
\nonumber
\end{eqnarray}
and the degree of superposition
\begin{eqnarray}
\varepsilon_A&=&a_1a_2+a_2a_3+a_1a_3, \\ \nonumber
\varepsilon_B&=&b_1b_2+b_2b_3+b_1b_3.
\end{eqnarray}
There are six kinds of entangled state according to the symmetry
between the branches within the entangled state. Here we only give
two examples, the other four are easy to be drawn:
\begin{eqnarray}
|\Psi_1\rangle &=& \frac{a_1b_1|1 1\rangle+a_2b_2|2
2\rangle+a_3b_3|3
3\rangle}{\sqrt{a_1^2b_1^2+a_2^2b_2^2+a_3^2b_3^2}}, \\
\nonumber |\Psi_2\rangle &=& \frac{a_1b_2|1 2\rangle+a_2b_1|2
1\rangle+a_3b_3|3
3\rangle}{\sqrt{a_1^2b_2^2+a_2^2b_1^2+a_3^2b_3^2}}. \\
\nonumber
\end{eqnarray}
The entanglements of these six states have the form
$\mathcal{E}_i=\alpha_i/\beta_i$. After some algebra, we can get
\begin{eqnarray}
\sum_i^6\frac{\alpha_i}{\mathcal{E}_i} &=& \sum_i^6\beta_i=2,
\\ \nonumber
\sum_i^6\alpha_i &=& 2\varepsilon_A\varepsilon_B.
\end{eqnarray}
For the maximal condition, when $\alpha_i=1/3$, $\beta_i=1/3$,
$\varepsilon_A=\varepsilon_B=1$, $\mathcal{E}_i=1$. For this
bi-qutrit entangled state, there is no definite form of reduced
entanglement.

For $4 \otimes 4$ system, there exist totally twenty-four different
states, each has four branches. The entanglement also can be written
as $\mathcal{E}_i=\alpha_i/\beta_i, i=1,2,...24$. We get the
following equations
\begin{eqnarray}
\sum_i^{24}\frac{\alpha_i}{\mathcal{E}_i} &=& \sum_i^{24}\beta_i=6,
\\ \nonumber
\sum_i^{24}\alpha_i &=& 4\varepsilon_A\varepsilon_B.
\end{eqnarray}
For the maximal condition, when $\alpha_i=3/8$, $\beta_i=1/4$,
$\varepsilon_A=\varepsilon_B=3/2$, $\mathcal{E}_i=3/2$.

Generally, for $n \otimes n$ system, there are $N=n!$ different
entangled states, set $\mathcal{E}_i=\alpha_i/\beta_i$, we can get
the general equations as
\begin{eqnarray}
\sum_i^N\frac{\alpha_i}{\mathcal{E}_i} &=& \sum_i^N\beta_i=(n-1)!,
\\ \nonumber
\sum_i^N\alpha_i &=& 2(n-2)!\varepsilon_A\varepsilon_B.
\end{eqnarray}
For the maximal condition, when $\alpha_i=(n-1)/2n$, $\beta_i=1/n$,
$\varepsilon_A=\varepsilon_B=(n-1)/2$, $\mathcal{E}_i=(n-1)/2$. That
is, for the two-party systems, when the number of states entangled
together increases, the Max($\mathcal{E}$) also increases, like the
superposition $\varepsilon$. This is reasonable since when more
states are entangled together, the amount of entanglement should
increase, and the state becomes more robust.

For the higher dimensional systems, the math is a little complex,
and the physical picture is not so clear. We do not analyze them
here. Next, we turn to the quantization of the four kinds of
correlations we defined in the last subsection.

For the entangled qudit, the entanglement is the shared coherence,
so the degree of entanglement $\mathcal{E}$ can be easily
calculated, according to the definition \ref{def:2} above. Also, as
we have discussed, we can employ different forms of entropy to
characterize the quantum information. At present, the well-known
quantities include von Neumann entropy, relative entropy, quantum
discord, squashed entanglement, entanglement of formation, cost, and
distillation, etc
\cite{Peres,Vidal02,bbps,Hill,Bennett,Vedral,Vidal99,Steiner,Ollivier,Luo,Christandl,Brandao,Gittsovich}.
We will study some of them in Appendix \ref{sec:MEASURE}.

For the decohered qudit, there is pure classical correlation without
shared quantum coherence. Also, there exist the complete projective
measurements, under which the classical density matrix remains
\cite{Ollivier,Luo}. The information is characterized by the von
Neumann entropy.

For the four kinds of ensemble state, see TABLE I, there exists
nonorthogonality. This kind of coherence is not globally shared by
the two parties $A$ and $B$, yet, it is shared by the local states
$|\psi_{\xi}^A\rangle$ of $A$, and the local states
$|\psi_{\xi}^B\rangle$ of $B$. The degree of nonorthogonality can be
defined as
\begin{equation}
\mathcal{\bar{O}}=\sum_{\mu=A, B}\sum_{\xi \neq \xi '}|\langle
\psi_{\xi}^{\mu}|\psi_{\xi '}^{\mu}\rangle|.
\end{equation}
In the sum, each element is smaller than one, yet, the sum can be
arbitrarily big even infinity. This is reasonable since the quantity
of coherence increases with the amount of the ensemble.
Particularly, we note that the separable state contains the quantum
nonorthogonality thus quantum information, although there is no
entanglement.

The coarse-grained classicality results from the classical parameter
$p_{\xi}$, so the classical information can be easily calculated via
the von Neumann entropy $S=-\sum_{\xi}p_{\xi} \log p_{\xi}$.

Last, we make some comments on the mixed state, which, in deed, is a
rather broad and rough method. The mixed state with purity smaller
than unity is a ``fragment'' of a global pure state with an unknown
part lost, thus, it is not easy to decide the exact structure of the
mixed state. At present, there exist lots of ways to detect the
entanglement, and different methods can be unified with the method
of entanglement witness in a way \cite{Brandao}. Yet, even we can
find the entanglement, we cannot decide the structure of the mixed
state. That is to say, to find how the mixed state is formed is more
valuable and practical. The six kinds of states studied above can
serve as some standard or assistant to detect the quantum
correlations. When a mixed state $\rho$ can be written as similar to
any standard state, labeled as $\sigma$, that is, if $\rho=\lambda
\sigma + (1-\lambda) \rho'$, then the quantum correlation
(entanglement, nonorthogonality, etc) of $\rho$ can be defined as
that of $\sigma$ times $\lambda$. $\rho'$ can be viewed as the
disturbance or perturbation to $\sigma$. This method is similar with
the relative entropy method \cite{Vidal99} and the
Lewenstein-Sanpera (LS) decomposition \cite{Sanpera}.

To sum up, in this section, we quantify the degree of superposition
and entanglement, and also nonorthogonality. Physically,
entanglement is directly based on superposition, and
nonorthogonality is also the result of superposition. The four kinds
of correlations can be easily calculated.

\section{CONCLUSION}
\label{sec:CONCLUSION}

The weirdness of QM is a result of the nature of quantum coherence,
the abstractness of quantum description, and the confusions on
quantum logic and philosophy. The concept of scope, which is exotic
for CM, Relativity, statistical mechanics, also classical wave
mechanics, just represents the quantum origin. It describes the
possible (potential) action region and the systematic and coherent
structure of movement of a certain object. Superposition and
entanglement are the natural results of scope with definite physical
meaning. Also, we discussed the properties of four kinds of quantum
correlations and the six kinds of quantum states, and we show that
entanglement is related yet different with entropy. Further, the
concept of scope brings new methods to the standard QM. One is that
the scope itself has certain geometric features, just similar to the
matter itself. This may indicate that the structure of the motion of
the matter is more crucial than the structure of the matter itself.
The other is that, rather than the information-theoretic framework,
there exists a new kind of space, i.e., tangnet, on which the
quantum structure with entanglement exists \cite{wang}. As a result,
the concept of scope lays the new starting point to the foundation
of QM. In the future, further efforts on, e.g., the geometrical
properties of tangnet, the comparison of scope with other
interpretations, also the relation between nonlocality and
entanglement, are needed.

In addition, the physical roles of entanglement in QIQC are also
widely concerned, we hope our approach can bring new ideas on some
tough problems, such as the bound entanglement \cite{bound}, the
analogy of entanglement to energy or entropy, etc.

\appendix

\section{Wave function of Ensemble system and Decoherence}
\label{sec:wfesd}

In Hilbert space, one state vector can be decomposed via orthogonal basis also non-orthogonal basis. Physically, the two decompositions can be endowed with different physical contents: for single system, there only exists orthogonal physical basis; while for ensemble, there can be non-orthogonal basis. For example, the spin state in one quantum dot can be up or down, or to the left or right in the diagonal basis; for a collection of spins in quantum dots, the superposed states of spins form the non-orthogonal basis. Mathematically, based on the measurement theory, especially the positive operator-valued measure (POVM) \cite{Nielsen}, we can see the WFES has the corresponding properties as the wave function of pure state.

For the pure state $|\phi\rangle=\sum_ic_i|i\rangle$, it satisfies: (1) Normalization: $\sum_i |c_i|^2=1$; (2) Completeness relation: $\sum_i |i\rangle\langle i|=1$; (3) Projective measurement: there exist a set of projective operators $\{\hat{P}_i=|i\rangle\langle i|\}$ satisfying $\hat{P}_i|\phi\rangle=c_i|i\rangle$, the expectation value  $\langle\hat{P}_i \rangle=|c_i|^2$; note that when the non-projective POVM operator acts on the pure state, it cannot distinguish its basis; (4) Probabilistic interpretation: $|c_i|^2$ means the probability to find state $|i\rangle$ with projector $\hat{P}_i$; (5) Observable: the expectation value of observable $\hat{A}$ is $\langle \hat{A}\rangle=\sum_i a_i |c_i|^2$, with $\hat{A}|i\rangle=a_i|i\rangle$; (6) Evolution: $|\phi\rangle$ also the basis $|i\rangle$ satisfy Schr\"odinger equation; (7) Decoherence: the non-diagonal elements in $|\phi\rangle\langle\phi|$ disappear, leading to state $\rho=\sum_i|c_i|^2|i\rangle\langle i|$, each part $|c_i|^2|i\rangle\langle i|$ forms the classical trajectory; (8) Quantum reference frame \cite{Rudolph}: to determine the preferred basis $\{|i\rangle\}$, the reference frame and certain interaction are needed to form the entangled state, i.e., the coherence within $|\phi\rangle$ delocalizes to form entanglement; thus, by ignoring the reference frame, the state $|\phi\rangle$ decohered.

For the WFES, the state in Eq. (\ref{eq:wfes}), the properties are similar while more complicated. We firstly note that if $k=1$, it reduces to the pure state case; if the states $\{ |\psi_k \rangle \}$ are orthogonal with each other, it goes to the totally classical (decohered) state. Instinctively, there exists one pure density matrix, labeled as $\varrho\equiv|\Psi\rangle\langle \Psi|$, which is
\begin{equation}
\varrho=\sum_{k,k'} \gamma_k \gamma_{k'}|\psi_k\rangle \langle \psi_{k'}|.
\end{equation}
To form the density matrix $\rho$, the non-diagonal elements disappear, with coefficients $\gamma_k \gamma_{k'\neq k}$, which is actually a kind of decoherence. The state $\varrho$ indicates that before the formation of the ensemble, there exists one decoherence process. Physically, the state $|\Psi\rangle$ stands for the ``field'' of the ensemble, and the ensemble emerges from the field after the decoherence process, i.e., the creation of particles \cite{Calzetta}. Note that here we do not study this issue with quantum field theory. We name this kind of coherence as {\em sub-coherence}, or coarse-grained coherence, and the decoherence as {\em sub-decoherence}. To arrive at the classical state, the standard decoherence is then needed. If we only account coherence by ignoring sub-coherence, we can view $|\Psi\rangle$ as the WFES, from which, the density matrix $\rho$ for the ensemble emerges. Further, the sub-decoherence can be introduced by one quantum reference frame $\sigma\equiv\sum_{k,k'} \gamma_k \gamma_{k'}|\phi_k\rangle\langle \phi_{k'}|$, which with $\varrho$ forms
\begin{equation}
\varpi=\sum_k |\gamma_k|^2|\psi_k\rangle \langle \psi_k|\otimes|\phi_k\rangle \langle \phi_k|,
\end{equation}
which is exactly the separable state \cite{Werner89}. Note that for pure state, the system and its reference frame form the entangled state.

Correspondingly, the basic properties of WFES are: (1) Normalization: $\sum_k |\gamma_k|^2=1$; (2) Completeness relation: $\sum_k |\psi_k\rangle\langle \psi_k|=1$;
(3) Measurement: there exist a set of POVM operators $\{\hat{E}_k=|\psi_k\rangle\langle \psi_k|\}$ satisfying $\hat{E}_k|\Psi\rangle=\gamma_k|\psi_k\rangle$, the expectation value $\langle\hat{E}_k \rangle=|\gamma_k|^2$; note that the measurement does not act on the fine-grained orthogonal basis directly; (4) Probabilistic interpretation: $|\gamma_k|^2$ means the probability to find state $|\psi_k\rangle$ with operator $\hat{E}_k$; (5) Observable: the expectation value of observable $\hat{A}$ is $\langle \hat{A}\rangle=\sum_k \alpha_k |\gamma_k|^2$, with $\hat{A}|\psi_k\rangle=\alpha_k|\psi_k\rangle$; (6) Evolution: states $|\Psi\rangle$ and $|\psi_k\rangle$ both satisfy Schr\"odinger equation, with which the Liouville equation for $\rho$ is equivalent; (7) Sub-decoherence: the non-diagonal elements in $|\Psi\rangle\langle\Psi|$ disappear, leading to quantum ensemble $\rho$; Decoherence: the party $|\psi_k\rangle$ each can further decohere, leading to classical state and classical trajectory, respectively; (8) Quantum reference frame: to determine the preferred basis $\{|\psi_k\rangle\}$, the reference frame $\sigma$ and certain interaction are needed to form the separable state, i.e., the sub-coherence within $|\Psi\rangle$ delocalizes to form separability; thus, by ignoring the reference frame, the state sub-decoheres; Further, if there exists the entanglement-type reference frame, the state eventually decoheres to classical trajectory.

Last, to illustrate the difference between coherence and sub-coherence, we give one simple example: suppose there are only two parties, $a$ and $b$, in the ensemble, and each party with only two orthogonal basis $|1\rangle$ and $|2\rangle$, and $|\psi_a\rangle=c_1^a|1\rangle+c_2^a|2\rangle$, $|\psi_b\rangle=c_1^b|1\rangle+c_2^b|2\rangle$. Note that even if party $b$ is not in the basis $|1,2\rangle$, there can be rotation which can rotate it to this basis, with effects absorbed in the coefficients. Thus, the density matrix of WFES is written as
\begin{widetext}
\begin{equation}
\varrho=\left( \begin{array}{cc} |\gamma_a|^2|c_1^a|^2+|\gamma_b|^2|c_1^b|^2+\gamma_a\gamma_b^*c_1^ac_1^{b*}+\gamma_b\gamma_a^*c_1^bc_1^{a*} & |\gamma_a|^2c_1^ac_2^{a*}+|\gamma_b|^2c_1^bc_2^{b*}+\gamma_a\gamma_b^*c_1^ac_2^{b*}+\gamma_b\gamma_a^*c_1^bc_2^{a*} \\
|\gamma_a|^2c_2^ac_1^{a*}+|\gamma_b|^2c_2^bc_1^{b*}+\gamma_a\gamma_b^*c_2^ac_1^{b*}+\gamma_b\gamma_a^*c_2^bc_1^{a*}& |\gamma_a|^2|c_2^a|^2+|\gamma_b|^2|c_2^b|^2+\gamma_a\gamma_b^*c_2^ac_2^{b*}+\gamma_b\gamma_a^*c_2^bc_2^{a*} \end{array} \right).
\end{equation}
\end{widetext}
The sub-coherence is represented as the coefficients $\gamma_a\gamma_b^*$ and its conjugate, when they disappear, the state $\varrho$ sub-decoheres to the quantum ensemble $\rho$. The coherence is represented as the coefficients $c_1^ac_2^{a*}$, $c_1^ac_2^{a*}$ and their conjugate, when they disappear, the state decoheres to the classical state, labeled as $\epsilon$, with diagonal elements $\epsilon_{11}=|\gamma_a|^2|c_1^a|^2+|\gamma_b|^2|c_1^b|^2$, $\epsilon_{22}=|\gamma_a|^2|c_2^a|^2+|\gamma_b|^2|c_2^b|^2$, and $\epsilon_{11}+\epsilon_{22}=1$.

\section{Scope and other methods}
\label{sec:SCOPES}

{\em Scope and QFT.} In QFT, the wave function is further viewed as
the operator, which is the well-known ``second quantization''. The
total number of particles is not conserved as there always exist the
fluctuations of field and creation/annihilatin of particles.
Particle itself is viewed as the excited state of field, and field
is viewed as a kind of matter, whose motion exists within itself.
Thus, the second quantization of wave function, i.e., scope, means
to view the structure of the motion as the equivalence of the
structure of matter, also to view the motion of the matter as the
same with the matter itself, which is reasonable for field. This is
also consistent with the mass-energy relation $E=mc^2$ from
Relativity. However, for particles, atoms etc in QM, the object
itself is different with its motion, thus, it is not proper to view
scope as a kind of matter; instead, scope should be viewed as a kind
of description of the logical structure of the motion of the object.

In history, another physical picture similar with QFT is mainly
developed by de Broglie, namely, the ``pilot wave'' approach
\cite{debroglie}, in which the particle with definite local form
sits at a certain place in the real pilot wave, and the particle can
be viewed as the high-energy concentrate singularity of the wave.
The pilot-wave describes the collective motion of two objects:
particle and wave, which indeed goes to the method of QFT. This
relates to the ``wave-particle duality'' that the particle can
perform both classical particle property and wave property, but not
both of them at the same time \cite{Bohm}. Also, the concept of
``matter wave'' was developed. However, according to the concept of
scope, the relation $\frac{m}{\lambda}=\frac{h}{c}$ should be
explained as that the object has the particle property ($m$) and
wave property ($\lambda$) within its own scope, consistent with the
wave-particle duality. As a result, the matter wave cannot be viewed
as a kind of matter, and indeed, the so-called matter wave does not
exist. We note that the other theory, Bohmian mechanics
\cite{bohm52,Dewdney}, in which the particle has the trajectory
restrained by the wave function, is consistent with the methods of
QTF and scope.

{\em Consistent History (CH)} \cite{ch}. This is a widely studied
form of QM, yet, there exist different approaches. This theory has
the similar character with our methods based on scope, particularly
the active operator. However, in CH, there is no reason for the
physical meaning of the decomposition of the identity. Next, we
present our understanding for the CH. Generally, CH takes the
evolution of state in time into account. The main quantity, the
decoherence functional \cite{ch} is defined as
\begin{equation}
\mathcal{D}(\vec{\alpha},
\vec{\alpha}')=\textrm{Tr}(\mathcal{C}_{\vec{\alpha}} \rho
\mathcal{C}^{\dagger}_{\vec{\alpha}'}),
\end{equation}
where $\rho$ is the density matrix. The class operator is
$\mathcal{C}_{\vec{\alpha}} \equiv \odot \prod_{\alpha=n}^1
P_{\alpha}^{i_{\alpha}}$, with $i_{\alpha}=\{1,2,..., n_{\alpha}\}$,
$\odot$ means that $P_{\alpha}^{i_{\alpha}}$ acts on the scope
instead of state. $P_{\alpha}^{i_{\alpha}}$ is the projective
decomposition of the identity,
\begin{eqnarray}
\sum_{i_{\alpha}=1}^{n_{\alpha}}P_{\alpha}^{i_{\alpha}}&=& 1, \\
\nonumber P_{\alpha}^{i_{\alpha}'} P_{\alpha}^{i_{\alpha}} &=&
\delta_{i_{\alpha}i_{\alpha}'}P_{\alpha}^{i_{\alpha}}.
\end{eqnarray}

When $\vec{\alpha}=\vec{\alpha}'$,
$\mathcal{D}(\vec{\alpha})=p_{\alpha}$, which is the probability of
the history, and the normalization $\sum_{\alpha}p_{\alpha}=1$.

When $\vec{\alpha}\neq \vec{\alpha}'$, if $\mathcal{D}(\vec{\alpha},
\vec{\alpha}'=0)$, then we say the history is decohered.

\begin{figure}
\includegraphics[width=0.33\textwidth,height=0.25\textwidth]{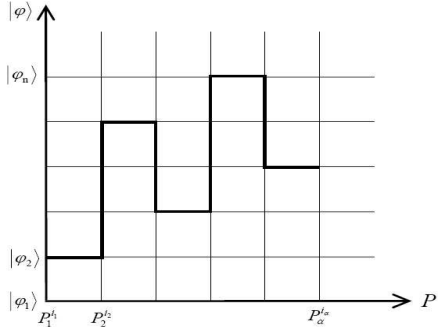}
\caption{The consistent history in the state-operator lattice space.
There exist $n$ eigenstates within the scope, and the number of the
projection (activer) $\alpha$ represents the evolve of time. When
the states become continuum, the lattice space becomes continuous
space. The black polylines stands for the consistent history.}
\label{fig.ch}
\end{figure}

In figure \ref{fig.ch}, we show one consistent history. We put state
and operator on the equal footing, then form the state-operator
space. $\{|\psi_i\rangle, i=1,2,...,n\}$ is the eigenstate set as
the ordinate, the abscissa is the $P_{\alpha}^{i_{\alpha}}$, where
$\alpha$ associates with time $t_{\alpha}$. For simplicity, we set
$n=6$, i.e., there are six states within the scope. From the figure,
we can easily get the activers of the history: $P_1^2, P_2^5, P_3^3,
P_4^6, P_5^4$.

In addition, there can be other kinds of history, e.g., several
activers can act together as
$\sum_{i_{\alpha}}P_{\alpha}^{i_{\alpha}}$ causing the superposed
state.

{\em Kraus Operator for open system} \cite{Kraus}. We study via the
widely concerned model of the pure universe \cite{wang,Popescu}. Let
universe $\mathcal{U}$ be composed with system $\mathcal{S}$ and
environment $E$, labeled as $\mathcal{U}=\mathcal{S}+E$. The initial
state is $|\psi_S(0)\rangle$ and $|\psi_E(0)\rangle$,
correspondingly, the initial state for the universe is thus
$|\psi(0)\rangle=|\psi_S(0)\rangle|\psi_E(0)\rangle$. The evolution
of the wave function is $|\psi(t)\rangle=U(t)|\psi(0)\rangle$, with
the evolution operator $U(t)$. The reduced density matrix for the
system is
\begin{eqnarray}
\rho_S(t)&=&\textrm{Tr}_E|\psi(t)\rangle \langle\psi(t)|
\\ \nonumber
&=&\sum_i\langle i| U(t)|\psi_S(0)\rangle|\psi_E(0)\rangle
\langle\psi_S(0)|\langle\psi_E(0)|U^{\dagger}(t)| i \rangle \\
\nonumber &=& \sum_iK_i(t)\rho_S(0)K^{\dagger}_i(t),
\end{eqnarray}
where $\rho_S(0)=|\psi_S(0)\rangle\langle\psi_S(0)|$,
$K_i(t)=\langle i |U(t)|0\rangle$, and we let
$|\psi_E(0)\rangle=|0\rangle$. The normalization rule is
\begin{equation}
\sum_i \langle 0 |U^{\dagger}(t)|i\rangle\langle i|U(t)|0\rangle=1.
\end{equation}

The Kraus operator representation demonstrates the openness and the
evolution in time of quantum system. However, although there is
evolution in time, the logical relations between operators remains.
For instance, the relation $\sum_iK_i^{\dagger}K_i=1$ does not
depends on time. Below, we show that this method is consistent with
the CH method.

The physical role of the evolution operator $U(t)$ can be
substituted by the class operator $\mathcal{C}_{\alpha}$, the state
of the system at time $t_n$ is
$|\psi_S(t_n)\rangle=\mathcal{C}_{\alpha}|\psi_S(0)\rangle$, where
$|\psi_S(0)\rangle$ is the initial state. We know
\begin{equation}
\mathcal{D}(\alpha)=\textrm{Tr}(\mathcal{C}_{\alpha}|\psi(0)\rangle
\langle\psi(0)| \mathcal{C}^{\dagger}_{\alpha}),
\end{equation}
and
\begin{eqnarray}
\sum_{\alpha}\mathcal{D}(\alpha) &=&
\sum_{\alpha}\textrm{Tr}\mathcal{C}_{\alpha}|\psi(0)\rangle
\langle\psi(0)| \mathcal{C}^{\dagger}_{\alpha} \\ \nonumber \\
\nonumber &=&
\textrm{Tr}\sum_{\alpha}\mathcal{C}_{\alpha}|\psi(0)\rangle
\langle\psi(0)| \mathcal{C}^{\dagger}_{\alpha} \\ \nonumber &=&
\textrm{Tr}\sum_{\alpha}\mathcal{C}_{\alpha}\rho_S(0)\mathcal{C}^{\dagger}_{\alpha}
\\ \nonumber &=& 1,
\end{eqnarray}
thus
$\sum_{\alpha}\mathcal{C}_{\alpha}\mathcal{C}_{\alpha}^{\dagger}=1$,
which is the same as the normalization of the Kraus operator,
physically.

{\em Interacting Faculties}. The concept of ``potentiality'',
originated from Aristotle \cite{Aristotle}, is quite widely studied
in QM. The actuality (reality) and potentiality were systematically
studied recently, where the authors stated that potentiality could
be a mode of existence, and the concept of ``faculty'' was
introduced to replace ``entity'' \cite{Ronde}. The faculty describes
a level which does not pertain to things but rather to potential
actions. Possessing a faculty offers the ability to do something.
According to our understanding, yet, the concept of faculty is the
analogy of potentiality. This approach relates to the property $4$
in section \ref{sec:PROPERTIES}, yet it does not aim to provide the
direct physical meaning of wave function and the superposition
principle.

\begin{figure}
\includegraphics[scale=0.33]{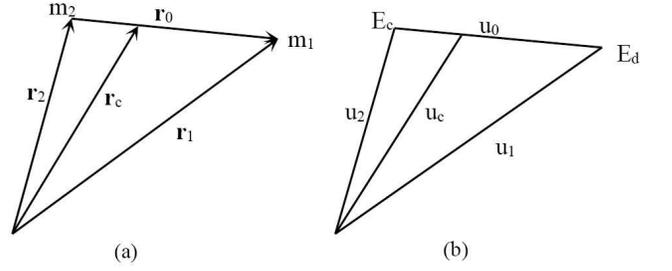}
\caption{The models for the two-body problem in CM (a) and the
mixtures of qubit in QM. The meaning for the symbols is explained in
the text.} \label{fig.red}
\end{figure}

\section{Reduced entanglement}
\label{sec:REDUCED}

Consider the mixtures of Bell states $|\Psi\rangle=(|01\rangle \pm
|10\rangle)/\sqrt{2}$, $|\Phi\rangle=(|00\rangle \pm
|11\rangle)/\sqrt{2}$. According to our definition, the state
$|\Psi\rangle$ is the ``cross'' state, and $|\Phi\rangle$ is the
``direct'' state. So the mixture (ensemble) can be viewed as the
mixture of two kinds of systems. The classical analogy is, e.g., the
mixture of two kinds of gases. There exists the center of mass for
each of the gases, so the problem can be simplified as the two-body
problem in classical mechanics, as shown in Fig. \ref{fig.red}(a).
The two objects $m_1$ and $m_2$ exist in the configuration space,
the coordinates are ${\bf r_1}$ and ${\bf r_2}$, respectively. The
kinetic energy is
\begin{eqnarray}
E_k &=& \frac{1}{2}(m_1 \dot{{\bf r}}_1^2+m_2 \dot{{\bf r}}_2^2) \\
\nonumber &=& \frac{1}{2}(M\dot{{\bf r}}_c^2+\mu\dot{{\bf r}}_0^2),
\end{eqnarray}
where $M=m_1+m_2$ is the total mass, $\mu= \frac{m_1m_2}{m_1+m_2}$
is the reduced mass, ${\bf r_c}=\frac{m_1{\bf r_1}+m_2{\bf
r_2}}{m_1+m_2}$ is the coordinate of the mass center of the system,
and ${\bf r_0}={\bf r_1}-{\bf r_2}$ is the relative displacement
between the two objects.

Now, consider velocity ${\bf v}=\dot{{\bf r}}$. As we know, there
exists the maximum speed as the light speed $c$. Let $c=1$, thus,
the set of speed $v=|{\bf v}|/c$ forms the probability space with
$0\leqslant v\leqslant1$. The kinetic energy is written as
\begin{equation}
E_k=\frac{1}{2}(M v_c^2+\mu v_0^2).
\end{equation}

Consider the mixtures of Bell states. The density matrix can be
written as
\begin{equation}
\rho=p_1 | \Phi \rangle \langle \Phi |+p_2 | \Psi \rangle \langle
\Psi |.
\end{equation}

The entanglement of the mixture is
\begin{eqnarray}
\mathcal{E}_{\rho} &=& p_1 \mathcal{E}_d+ p_2 \mathcal{E}_c \\
\nonumber &=& (\mathcal{E}_d+ \mathcal{E}_c)p_c+\frac{\mathcal{E}_d
\mathcal{E}_c}{\mathcal{E}_d+\mathcal{E}_c}p_0,
\end{eqnarray}
where by introducing the probability amplitude $u=\sqrt{p}$,
$u_c=\frac{u_1\mathcal{E}_d+u_2\mathcal{E}_c}{\mathcal{E}_d+\mathcal{E}_c}$,
$u_0=|u_1-u_2|$, and $\frac{\mathcal{E}_d
\mathcal{E}_c}{\mathcal{E}_d+\mathcal{E}_c} \equiv
\mathcal{E}^{\dagger}$ is the reduced entanglement. The situation is
depicted in Fig. \ref{fig.red}(b).

By comparison, the entanglement is the analogy of mass, the
probability is the analogy of $v^2$.

For high-dimensional and many-body problems, there is no obvious
form of reduced mass, also the reduced entanglement. Yet, we can
apply the rough bi-party partition to deduce the apparent reduced
mass or entanglement.

However, we have to say, there are significant differences. For
instance, the quantum probabilities $p_1$ and $p_2$ each exist in
the probability space, and also they together exist in another
probability space. But, the classical speed (square) $v_1^2$ and
$v_2^2$ do not exist in the probability space with each other. This
relates to the problem of Special Relativity, this simple mode may
have implication for the relation between QM and Relativity.

\section{Entanglement measure}
\label{sec:MEASURE}

According to our study, entanglement and quantum information
(entropy) are different. Entanglement is the shared coherence by the
parties of the entangled state. Also, we find nonorthogonality is
the shared coherence within the local states of each party of the
whole state. Quantum information ($\mathcal{I}_q$) is the improved
form of classical information (correlation) ($\mathcal{I}_c$) due to
nonorthogonality ($\bar{\mathcal{O}}$) or entanglement
($\mathcal{E}$). Thus, $\mathcal{I}_q$ can be the function of
$\bar{\mathcal{O}}$ or entanglement $\mathcal{E}$. Here, we compare
our definition with the usual entanglement measure.

{\em Entanglement of formation.} In QIQC, the mixtures of Bell
states, i.e., ensemble of qubits, are often prepared, the
entanglement of formation, also cost and distillation are studied
\cite{bbps,Hill,Bennett}. The entanglement of formation, which
measures quantum information instead of entanglement, is defined as
\begin{equation}
E_f(\rho)=\textrm{min}\sum_i^n p_i E(\psi_i),
\end{equation}
and the density matrix is written as $\rho=\sum_i^n p_i
|\psi_i\rangle \langle \psi_i|$. This decomposition is the ensemble
of entangled qudit. Here we prove one simple property of $E_f(\rho)$
as follows.

When the density matrixes $\omega$, $\sigma$, and $\rho$ satisfying
\begin{eqnarray}
\omega&=&r_1 \rho+r_2 \sigma \\ \nonumber &=& r_1 \sum_{i=1}^{i=n}
p_i |\psi_i\rangle \langle \psi_i|+r_2  \sum_{j=1}^{j=m} q_j
|\phi_j\rangle \langle \phi_j|
\\ \nonumber &=& \sum_{i=1}^{i=n} r_1 p_i |\psi_i\rangle \langle
\psi_i|+\sum_{i=n+1}^{i=n+m} r_2 q_i |\phi_i\rangle \langle \phi_i|.
\end{eqnarray}
The entanglement of formation of $\omega$ is
\begin{eqnarray}
\nonumber E_f(\omega)&=& \textrm{min} (r_1 \sum_{i=1}^{i=n} p_i
E(\psi_i) + r_2 \sum_{i=n+1}^{i=n+m} q_i E(\phi_i)) \\ &=&r_1
E_f(\rho)+ r_2 E_f(\sigma).
\end{eqnarray}

{\em Concurrence.} Concurrence relies on the permutation symmetry of
entangled state \cite{Werner89}. For pure state, concurrence
describes the coherence. There are three kinds of basis to express
the qubit.

(a). ``Magic basis'' $\{ |e_i\rangle \}$ \cite{Hill}. The qubit is
written as
$|\psi\rangle=\sum_i\alpha_i|e_i\rangle=(\alpha_1+i\alpha_2, i
\alpha_3+\alpha_4, i \alpha_3-\alpha_4, \alpha_1-i \alpha_2)^T/2.$
Concurrence is
\begin{equation}
  C=|\sum_i \alpha_i^2|=\alpha_1^2+\alpha_2^2+\alpha_3^2+\alpha_4^2.
\end{equation}

(b). ``Computation basis'' $|00\rangle$, $|01\rangle$, $|10\rangle$,
$|11\rangle$. The qubit is written as $|\psi\rangle=(a,b,c,d)^T$,
and concurrence is
\begin{equation}
  C= 2|ad-bc|.
\end{equation}
It is easy to see that the basis in case (a) can be viewed as the
special case of (b).

(c). Schmidt basis $|ii\rangle$.  The qubit is written as
$|\psi\rangle=x|00\rangle+y|11\rangle$, with $x^2+y^2=1$, the
density matrix is
\begin{equation}
\rho = \left( \begin{array}{cccc}x^2 & 0&0& xy\\ 0& 0& 0& 0\\ 0 &0& 0& 0\\
xy& 0& 0& y^2\end{array} \right),
\end{equation}
and concurrence is
\begin{equation}
C= 2|xy|.
\end{equation}
The Schmidt basis can be rotated to the ``computation basis'' in
case (b) by the local unitary transformation $U_A\otimes U_B$
\cite{Abouraddy}.

The above three kinds of basis give the same results of concurrence.
The entanglement of formation is defined as
$E_f=h(\frac{1+\sqrt{1-C^2}}{2})$, where the binary entropy function
$h(x)=-x\log_2 x-(1-x)\log_2 (1-x)$. We can form the
entropy-concurrence matrix ($\Lambda$) as follows
\begin{equation}
\Lambda= \frac{1}{2}\left( \begin{array}{cc} 1+\sqrt{1-C^2} &  C\\
C& 1-\sqrt{1-C^2}\end{array} \right).
\end{equation}
Then, it is obvious that the entropy $E_f$ describes the property of
the diagonal elements (population), and the concurrence $C$
describes the non-diagonal elements (coherence).

{\em Negativity and Robustness.} For pure state, the definition of
negativity \cite{Vidal02} is the same with our definition of the
degree of entanglement. Negativity is usually viewed as the absolute
value of the sum of the negative eigenvalues of $\rho^{T_A}$
\begin{equation}
\mathcal{N}= \frac{||\rho^{T_A}||_1-1}{2}=|\sum_i \mu_i|,
\end{equation}
where $\mu_i$ is the negative eigenvalues.

In deed, the negative eigenvalues are the results of the shared
coherence of the entangled state. For pure qubit, the negativity is
the same with concurrence, that is to say, the definition of
negativity is more general than concurrence for bi-party pure state.

The robustness \cite{Vidal99} is the twice of the negativity for
pure state. Physically, it describes the robustness of the
entanglement to the minimal amount of external classical noise. The
role of noise is to disturb and destroy the coherence. Physically,
the robustness can describe the shared coherence, thus give the
right quantity of entanglement. However, it is not easy to calculate
since it is the indirect way to measure entanglement.

In addition, we discuss a little of the physical meaning of
concurrence and negativity for the mixed state condition. It is
well-known that the two are often different for mixed state, though
the same for the pure state case. The physical reason is quite
clear: they are the results of two different entanglement witness
(operator) \cite{Brandao}. Physically, in the mixed state there can
be nonorthogonality, which can influence concurrence and negativity
without being detected. That is to say, it is not proper to
generalize concurrence and negativity to the mixed state, since the
result is not necessarily entanglement.

{\em Relative entropy.} The quantum relative entropy of entanglement
\cite{Vedral} is defined as

\begin{eqnarray}
E_r(\rho)&=&  \inf_{\sigma\in X} S(\rho||\sigma) \\ \nonumber &=&
\inf_{\sigma\in X} \textrm{tr} (\rho \log \rho-\rho \log \sigma),
\end{eqnarray}
where the set $X$ can be taken as the set of separable states,
states with positive partial transpose, nondistillable states, etc.
The quantum discord \cite{Ollivier,Luo} and squashed entanglement
\cite{Christandl}, which we have studied in Ref. \cite{wang}, can be
viewed as the relative entropy.

Physically, $E_r$ does not quantify entanglement directly, yet it
relates to entanglement. The relation between $E_r$ and entanglement
also nonorthogonality can not be deduced easily especially for mixed
state, since the detailed structure of the mixed state and the
distribution of coherence are not clear. As a result, the relative
entropy is only an effective way to measure quantum information
instead of entanglement.

\end{document}